# Semi-Quantitative Analysis and Seroepidemiological Evidence of Past Dengue Virus Infection among HIV-infected patients in Onitsha, Anambra State, Nigeria

[1]Nwachukwu, D., [1]Oketah, E. N., [2]Ugwu, C. H., [1]Innocent-Adiele, H. C., [1]Adim, C. C., [3]Onu, E.N., [1]Chukwu, A. O., [5]Nwankwo, G. A., [1]Igwe, M. U., [1]Okerentugba, P. O. & *[1]Okonko, I. O.

[1]Virus & Genomics Research Unit, Department of Microbiology, University of Port Harcourt, Port Harcourt, Nigeria. *Corresponding author's e-mail address: iheanyi.okonko@uniport.edu.ng; Tel: +2347069697309, ORCID iD: 0000-0002-3053-253X
[2]Department of Applied Microbiology, Nnamdi Azikiwe University, Awka, Anambra State, Nigeria.
[3]Department of Medical Microbiology, Faculty of Basic Clinical Medicine, Alex-Ekueme Federal University, Ndufu-Alike, Ikwo, Ebonyi State, Nigeria.
[4]Immunology & Vaccinology Unit, Department of Pharmaceutical Microbiology, University of Port Harcourt, Port Harcourt, Nigeria
[5]Department of Public Health, Madonna University Nigeria, Rivers State, Nigeria.

## ABSTRACT

Despite its endemic nature as well as the recent outbreaks, information on the opportunistic DENV in Anambra state has been sparse. This study thus aimed to give seroepidemiological evidence of past dengue virus infection among HIV-infected patients in Onitsha, Anambra State, Nigeria. Plasma from 94 HIV-infected patients who were attending Saint Charles Borromeo Hospital, Onitsha in Anambra State, Nigeria was tested for IgG antibodies specific to the dengue virus by IgG ELISA assay. The prevalence of past dengue virus infection was 61.7% (n = 58/94). This study showed age group 0-15 years (77.30%), female gender (65.1%), married (63.9%) and no formal level (100.0 %) as the highest seropositivity among the study participants. In terms of immunological and virological markers, greater IgG seroprevalence was observed in individuals with a viral load of <40 copies/ml (64.0%) and a CD4 count of >350 cells/µl (63.2%). The high IgG seropositivity of Dengue Virus (DENV) among HIV-infected individuals on Onitsha is cause for concern.

## 1. INTRODUCTION

Dengue fever is a mosquito-borne viral illness that affects millions of people worldwide, particularly in tropical and subtropical regions. Dengue fever is caused by the dengue virus, which belongs to the Flaviviridae family. There are four distinct serotypes (DENV-1, DENV-2, DENV-3, and DENV-4) that can cause the disease. Dengue virus is primarily transmitted to humans through the bite of infected Aedes mosquitoes, particularly *Aedes aegypti* and *Aedes albopictus* (Albasheer et al., 2021; Dolan et al, 2021; Eltom et al., 2021). These mosquitoes are commonly found in urban and semi-urban areas.

Dengue fever can manifest in various forms, ranging from asymptomatic or mild flu-like symptoms to severe, potentially life-threatening conditions. HIV, or human immunodeficiency virus, is a global health challenge that affects millions of people. HIV is a retrovirus that attacks the immune system, specifically CD4 T cells, weakening the body's ability to fight infections. It is primarily transmitted through unprotected sexual intercourse, sharing of contaminated needles, and from mother to child during childbirth or breastfeeding. It can also spread through blood transfusions or organ transplants and occupational exposure to infected blood (UNAIDS, 2021; CDC, 2022; WHO, 2023).

Coinfection with these two significant viruses, dengue virus (DENV) and human immunodeficiency virus (HIV), presents unique challenges and health concerns. Areas with high HIV prevalence often overlap with regions where DENV is endemic. Coinfection cases, where individuals are infected with both DENV and HIV, are reported in these regions. DENV-HIV coinfection can present unique clinical challenges. While DENV may exacerbate HIV replication and disease progression, HIV-related immunosuppression can lead to more severe dengue symptoms (Pang et al., 2015; Ayukekbong, 2015; Osarumwense et al., 2022).

This study thus aimed to give seroepidemiological evidence of past dengue virus infection among HIV-infected patients in Onitsha, Anambra State, Nigeria.

## 2. MATERIALS AND METHODS

### 2.1. Study Area and Design
The focus of this investigation centred on Anambra State, positioned in the east-central region of Nigeria, specifically, the investigation was conducted within Saint Charles Borromeo Hospital in Onitsha. This research utilized a cross-sectional approach, encompassing the collection and examination of samples using a Dengue virus IgG ELISA kit.

### 2.2. Sample Size Estimation
The sample size for this study was calculated using the established formula (MacFarlane, 1997; Niang et al., 2006; Awando et al., 2013). $N = [Z^2(pq)] / d^2$. The total sample size was calculated as 40.

### 2.3. Population for the Study
Ninety-four individuals who were HIV-infected from the study population were chosen randomly and enlisted in the study. Approximately, 94 confirmed HIV-positive patient samples were collected and stored in a sterile manner (labelled 1-94) from Saint Charles Borromeo Hospital located in Onitsha within the state of Anambra, Nigeria, facilitated by trained medical laboratory professionals. These samples were later transported to the Virus & Genomics Research Unit within the Department of Microbiology at the University of Port Harcourt, where the laboratory analyses were conducted. Alongside the testing, relevant clinical history, behavioural patterns, and demographic information were also gathered.

### 2.4. Sample and Sampling Techniques
The technique of stratified sampling was employed to select sample specimens. Each eligible patient received a unique number, which was recorded on the patient's card, and assisted by the laboratory assistants or hospital staff. Alongside the blood samples, pertinent demographic data, behavioural and lifestyle patterns, and clinical history were collected. The gathered demographic information was categorized into several groups: sexes (males, M, and females, F); age groups (0-15 years, 16-30 years, 31-45 years, 46-60 years, and 61-75 years); highest educational achievement; and marital status (single, married). Clinical history was also assessed (Table 1).

### 2.5. Sample Collection
Venipuncture was utilized to collect five millilitres (5 ml) of venous blood from each participant. The gathered blood was placed in EDTA bottles and then subjected to centrifugation at 3000 rpm for five minutes. This process aimed to separate the plasma required for detecting Dengue Virus antibodies. The obtained plasma was stored at -20°C before undergoing laboratory testing. The ELISA technique was employed using kits designed for capturing IgG antibodies to facilitate the testing process. The data consisted of both quantitative and qualitative information from study participants within the Saint Charles Borromeo Hospital in Onitsha.

**2.6. Serological Analysis**
Sample testing was conducted using ELISA techniques, following the manufacturer's guidelines for Dengue virus IgG tests. The samples consisted of plasma (EDTA). Results were obtained and interpreted using an ELISA reader (ELx800) at 450nm and 630nm wavelengths. The analysis of the data was conducted using the Chi-square test. Descriptive statistics were used to present demographic summary tables, which included factors such as age range and gender.

**Data Analysis**
Information about socio-demographic attributes, including age, gender, marital status, and educational status, along with clinical characteristics (Viral load and CD4 count) was collected from the HIV patient registration book at the hospital. After undergoing validation, this data was entered into the WPS spreadsheet version 12.2 for further evaluation. Initially, patients' data were categorized into groups based on their relative distribution for each characteristic. Subsequently, the data was grouped to reflect overall IgG prevalence. The variables were evaluated using the chi-square statistical analytic test, and the data were analyzed using WPS version 12.2.

**3. RESULTS**
**3.1. Socio-demographical and Clinical Characteristics of Study Participants**
The categorical data of the study participants' socio-demographical and clinical characteristics were appropriately organized as can be seen in Tables 3 and 4 respectively

**3.2. Overall IgG Assay Results**
As shown in Table 3, this study reported a 61.7% seropositivity for IgG

**3.3. DENV IgG Seroprevalence Concerning Socio-demographical Characteristics**.
In terms of age, the highest IgG seropositivity was seen in people 0-15 years old (77.3%), while the lowest was seen in people 16 to 30 years old (47.1%) (Table 1). Among gender, females had the highest IgG seropositivity at 65.1%, whereas males were at 54.8% (Table 1). Married participants displayed the highest IgG seropositivity rates (63.9%), in contrast to single participants (60.3%) as shown in Table 1. Regarding educational status, IgG antibody demonstrated 100.0% seropositivity among those without any formal education, while those with tertiary education had the lowest seropositivity (55.9%) (Table 1).

**3.4. DENV IgG Seroprevalence Concerning Clinical Characteristics**
Regarding the viral load, individuals with PVL of <40 copies/ml exhibited the highest IgG seropositivity (64.0%), while those with PVL between 40-1000 copies/ml displayed the lowest (57.1%) as portrayed in Table 1. For the CD4 count, study participants above 350 cells/μl were observed to have the highest IgG seropositivity at 63.2% while those with 200-350 cells/μl showed the lowest seropositivity of 58.6% (Table 1).

**3.5. Statistical Analysis**
Age (p-value = 0.46), gender (p-value = 0.34), marital status (p-value = 0.73),

education (p-value = 0.26), viral load (p-value = 0.95) and CD4 count (p-value = 0.92) were not significant risk factors among the study participants (Table 1).

**Table 1: IgG Results with Socio-demographic and Clinical Characteristics of Study Participants**

| Variables | Categories | No. Tested (%) | No. Positive (%) | $X^2_i$ | |
|---|---|---|---|---|---|
| **Age Group (Years)** | 0-15 | 9 (9.6) | 7 (77.8) | 0.98 | 0.46 |
| | 16-30 | 17 (18.1) | 8 (47.1) | 1.54 | |
| | 31-45 | 43 (45.7) | 29 (67.4) | 0.60 | |
| | 46-60 | 22 (23.4) | 12 (54.5) | 0.48 | |
| | 61-75 | 3 (3.2) | 2 (66.7) | 0.03 | |
| **Gender** | Male | 31 (33.0) | 17 (54.8) | 0.62 | 0.34 |
| | Female | 63 (67.0) | 41 (65.1) | 0.30 | |
| **Marital Status** | Single | 58 (61.7) | 35 (60.3) | 0.00 | 0.73 |
| | Married | 36 (38.3) | 23 (63.9) | 0.10 | |
| **Educational Status** | Primary | 21 (22.3) | 7 (87.5) | 2.25 | 0.26 |
| | Secondary | 49 (52.1) | 30 (60.0) | 0.06 | |
| | Tertiary | 22 (23.4) | 19 (55.9) | 0.49 | |
| | None | 2 (2.1) | 2 (100.0) | 1.24 | |
| **Viral Load (copies/ml)** | TND | 22 (23.4) | 14 (63.6) | 0.03 | 0.95 |
| | <40 | 25 (26.6) | 16 (64.0) | 0.06 | |
| | 40-1000 | 28 (29.8) | 16 (57.1) | 0.25 | |
| | >1000 | 19 (20.2) | 12 (63.2) | 0.02 | |
| **CD4 Count (cell/µl)** | <200 | 27 (28.7) | 17 (63.0) | 0.02 | 0.92 |
| | 200-350 | 29 (30.9) | 17 (58.6) | 0.12 | |
| | >350 | 38 (40.4) | 24 (63.2) | 0.03 | |
| **Total** | | **94 (100.0)** | **58 (61.7)** | | |

## 4. DISCUSSION

In this study, the prevalence of past dengue infection as indicated by IgG antibodies was 61.7%. Notably, this prevalence rate of 61.7% for IgG in this study exceeds the rates reported in previous studies conducted by Omatola et al. (2020) in Kogi State, Nigeria (20.5%) and Emeribe et al. (2021) in Nigeria (34.7%). Furthermore, it differs from the rate of 50.52% reported in a previous study in Ilorin, Kwara State by Kolawole et al. (2017), the 51.0% (Isaac, 2022) and 2.1% (Okonko et al., 2023) reported previously in Port Harcourt, Nigeria.

The finding of this study is lower than the 77.0% prevalence reported by Adeleke et al. (2016) in Osogbo, Southwestern Nigeria, as well as the 77.1% prevalence in Southeast Nigeria reported by Emeribe et al. (2021). Additionally, Ayukekbong (2015) reported a prevalence of 73.0% in Oyo State, specifically within the city of Ibadan. These varying prevalence rates can be attributed to several factors affecting

DENV distribution, including environmental elements (such as temperature, rainfall, weather, and vegetation), population density, quality of life, poverty, and personal hygiene (Ooi & Gubler, 2009a,b; Seposo et al., 2023).

The study revealed that the highest prevalence of dengue fever IgG, concerning age, was observed among individuals in the age group of 0-15 years, surpassing the prevalence rates among the other age groups. This finding is not consistent with Mustapha et al. (2017), who reported a higher seropositivity rate among individuals above 50 years in FCT, Abuja, Nigeria. Notably, high DENV seropositivity was identified across all age groups. The 77.8% prevalence among those aged 0-15 years and the 66.7% prevalence in the 61-75 years age group are likely attributable to the relatively underdeveloped and declining immune systems of these respective age groups. Their susceptibility is further compounded by their exposure to *Aedes spp* mosquitoes, which are highly prevalent in Anambra. The increased seropositivity observed in the other age groups, especially the 31-40 years can be explained by their participation in outdoor activities and work.

This observation of the highest prevalence occurring among the 0-15 years age group contradicts the findings of Mwanyika et al. (2021), who reported that age was a risk factor, associating prevalence with old age. It also deviated from that of Okonko et al. (2023), who reported a higher prevalence among 16-20 years. On the other hand, Tsheten et al. (2020) reported that children were more susceptible. In this study, children were closely followed by the elderly in terms of IgG prevalence, indicating that the virus finds it easier to infect these age groups due to the status of their immune systems.

Interestingly, a higher prevalence of dengue fever IgG was observed among females than males in this study. This finding is in contrast with the observations made by Omatola et al. (2021), who reported a higher seropositivity among males in Anyigba, Kogi State, Nigeria. It is also in opposition with Ugwu et al. (2018), who reported a higher prevalence of IgG in males in Jos, Plateau State. However, Mustapha et al. (2017) reported a higher prevalence among females in FCT in agreement with this study. Outdoor activities and socio-economic differences may account for the increased female IgG prevalence.

The results of this study indicated a higher prevalence of dengue IgG among married individuals compared to those who were single. This finding is consistent with the observations of Omatola et al. (2021), who reported a positive link between marriage and IgG seropositivity. Additionally, this finding aligns with the conclusions of Emeribe et al. (2021), who also established that married participants exhibited a significantly higher prevalence of DENV. The increased dengue prevalence among married participants in this study may be attributed to the proximity often maintained within marital households. Consequently, DENV-carrying mosquitoes that bite one spouse are highly likely to bite the other (Omatola et al., 2021). However, the present observation deviated from that of Okonko et al. (2023), who reported a higher prevalence among singles.

The findings of this study demonstrated a higher prevalence of dengue virus IgG among individuals with no formal education and primary education compared to those with secondary and tertiary education. Our results were not consistent with the observations made by Omatola et al. (2021), who reported a positive correlation between formal education and higher IgG seropositivity. However, the present observation aligned with that of Okonko et al. (2023), who reported a higher prevalence among primary education holders.

For clinical characteristics, it was observed that participants with a viral load of <40 copies/ml exhibited the highest IgG prevalence. This study did not find a significant association between HIV viral load and the seropositivity of dengue virus IgG. However, this present observation deviated from that of Isaac's (2022) study, which reported a higher prevalence among participants with a viral load of >1000 copies/ml in Port Harcourt, Nigeria.

Furthermore, a higher prevalence of dengue fever IgG was noted among individuals with CD4 counts < 200 cells/µl and those with CD4 counts >350 cells/µl compared to those with CD4 counts between 200-350 cells/µl. This present observation is consistent with that of Isaac's (2022) study, which reported a higher prevalence among participants with CD4 counts < 200 cells/µl in Port Harcourt, Nigeria. Importantly, this study found no significant relationship between CD4 count and DENV IgG seropositivity. This finding is consistent with the results reported by Joob & Wiwanitkit (2014), who observed a similar lack of association between CD4 count and DENV IgG seropositivity.

## 5. CONCLUSION

The results indicated that dengue virus (DENV) IgG prevalence was present among this group of people living with HIV in Onitsha, Anambra State, Nigeria. The elevated prevalence strongly suggests a significant presence of Aedes mosquitoes carrying DENV within the population. The high IgG seropositivity of DENV among residents of Onitsha is cause for concern. People living in or visiting Onitsha should proactively adopt preventive measures to mitigate the risk of DENV infection. These may include measures to minimize mosquito exposure, such as the use of insect repellents, wearing protective clothing, and implementing mosquito control measures in and around living spaces. Public health awareness and education campaigns may also play a crucial role in promoting community-wide preventive practices.

**Compliance with ethical standards**
**Acknowledgement**
The Saint Charles Borromeo Specialist Hospital's administration in Onitsha, Anambra State, Nigeria, and everyone who consented to participate in the study is appreciated by the authors for their approval.

*Disclosure of conflict of interest*
There are no competing interests, according to the writers.

*Statement of ethical approval*
Every author certifies that the University of Port Harcourt Research Ethics Committee reviewed and approved every experiment. As a result, the 1964 Declaration of Helsinki's ethical guidelines are adhered to throughout the study.

*Statement of informed consent*
"All authors affirm that every individual participant in the study gave their informed consent."